\documentclass[conference]{IEEEtran}
\IEEEoverridecommandlockouts

\usepackage{cite}
\usepackage{subcaption}
\usepackage{amsmath,amssymb,amsfonts}
\usepackage{algorithmic}
\usepackage{graphicx}
\usepackage{textcomp}
\usepackage{xcolor}
\usepackage{tikz}
\usepackage{microtype}
\usepackage{todonotes}
\usepackage{hyperref}

\definecolor{myblue}{HTML}{0070C0}
\DeclareRobustCommand{\circledred}[1]{%
  \tikz[baseline=(char.base)]{
    \node[shape=circle, fill=myblue, text=white, inner sep=1pt] (char) {{#1}};
  }%
}

\def\BibTeX{{\rm B\kern-.05em{\sc i\kern-.025em b}\kern-.08em
    T\kern-.1667em\lower.7ex\hbox{E}\kern-.125emX}}

\begin{document}

\title{Hazel Deriver: A Live Editor for Constructing Rule-Based Derivations}

\author{\IEEEauthorblockN{Zhiyao Zhong}
\IEEEauthorblockA{\textit{Computer Science and Engineering} \\
\textit{University of Michigan}\\
Ann Arbor, MI, USA \\
zhzy@umich.edu}
\and
\IEEEauthorblockN{Cyrus Omar}
\IEEEauthorblockA{\textit{Computer Science and Engineering} \\
\textit{University of Michigan}\\
Ann Arbor, MI, USA \\
comar@umich.edu}
}

\maketitle

\begin{abstract}
Students in programming languages and formal logic courses often struggle with constructing rule-based derivation trees due to the complexity of applying inference rules, the lack of immediate feedback, and the manual effort required for handwritten proofs. We present Hazel Deriver, a live, web-based editor designed to scaffold derivation construction through multiple layers of support. Built on the Hazel live programming environment, it provides a structured, interactive experience that encourages iterative exploration and real-time feedback. A preliminary user study with former students suggests that Hazel Deriver reduces the perceived difficulty of derivation tasks while improving conceptual understanding and engagement. We discuss the design of its layered scaffolding features and raise questions about balancing system guidance with learner autonomy.
\end{abstract}

\section{Introduction}

Rule-based derivations are an important part of computer science education. In courses on programming languages and formal logic, for example, students often need to build derivation trees that show how formal rules can be used to derive specific judgments, such as those specifying type checking, evaluation, or logical reasoning.

Despite their educational value for understanding formal systems, many students find derivations difficult, tedious, or frustrating. Students must apply rules recursively, draw large trees by hand, and carefully manage contexts and assumptions. Beginners often get stuck choosing the right rule or figuring out how to complete a step. Since these are typically done on paper and graded later, students often receive little or no feedback during the process\cite{bull1983derivation}. As a result, simple mistakes go unnoticed, and students miss chances to improve. Grading is also hard for instructors, who must check each step carefully. In large or online classes, these challenges make it difficult to scale support for derivation assignments.

To address these issues, we developed Hazel Deriver, a structured, interactive editor for constructing derivation trees. It is built on the Hazel live programming environment\cite{omar2019live}, which supports typed structure editing and continuous feedback. Hazel Deriver aims to help students explore derivations incrementally and receive immediate guidance during construction, without automating the derivation process itself.

The system supports learning through multiple layers of scaffolding. \textbf{Live Feedback.} Hazel Deriver checks each edit in real time and highlights incorrect steps (unless configured not to do so for pedagogical purposes), so students can revise their work without starting over. \textbf{Sidebar Documentation.} Rules are linked to synchronized documentation\cite{potter2022contextualized} that shows how abstract patterns map to user input. \textbf{Error Localization.} When an error is detected, the system highlights the relevant subterm to help students identify the issue.

The system inherits typed-hole mechanisms from the Hazel programming environment\cite{omar2019live}, which allows students to leave parts of the tree incomplete while working on other areas. This encourages a flexible, exploratory workflow.

Our tool builds on insights from systems like SaSyLF\cite{aldrich2008sasylf} and Holbert\cite{o2022holbert}, but differs by combining live structure editing, immediate feedback, typed holes, layered instructions, and contextualized documentation. Its design also aims to avoid excessive automation -- there is, by design, no proof search procedure in Hazel Deriver. Instead, the system provides feedback after students make key decisions.

This paper presents the design and reports findings from a user study with a small group of students who had prior experience with rule-based derivations on paper. We investigate the following research questions:

\textbf{RQ1:} How does Hazel Deriver affect the perceived difficulty of derivation problems and student confidence when completing derivation tasks?

\textbf{RQ2:} How do students interact with the system's layered scaffolding features?

Our study suggests that Hazel Deriver reduces perceived difficulty and frustration while helping students engage more confidently with derivation tasks.

\section{Related Work}

Hazel Deriver builds on a range of ideas developed to help students learn programming languages and formal reasoning. These include traditional proof assistants, educational editors, and structural programming environments.

Traditional proof assistants like Coq, Isabelle, and Agda are powerful tools for formal verification, but they are not well suited for beginners. These systems often require users to write complex scripts, understand abstract formal languages, and work in text-based interfaces. As noted by Keenan\cite{keenan2024learner}, such tools are mainly built for experts and are not designed to support early-stage learning or usability in educational settings.

Some tools have been created specifically for teaching. One example is SaSyLF\cite{aldrich2008sasylf}, which lets students write derivations in a rule-based format using a simple textual syntax. It checks proofs against user-defined systems and is often used in programming languages courses. However, it does not provide interactive feedback, visual structure, or contextual aids. Hazel Deriver improves on this by supporting real-time verification, structured editing, and integrated documentation.

Other systems have explored ways to make proof construction more accessible. Proof Pad\cite{eggensperger2013proof} offers a graphical interface for natural deduction proofs. Holbert\cite{o2022holbert} similarly supports visual proof construction and emphasizes user-friendly interaction for logic learning. Staudacher et al.\cite{staudacher2018conception} proposed an interactive learning environment for propositional logic that focuses on guided feedback and intuitive design. Mendes et al.\cite{mendes2014structure} introduced a system that supports structure-preserving manipulation of algebraic expressions. While these tools improve user experience, most are limited to specific domains such as propositional logic or algebra, and do not support systems like type inference or evaluation semantics. Moreover, only a few of these systems incorporate layered instructional scaffolding or live structural validation, which are central to Hazel Deriver's design.

Another related tool is RISCAL\cite{schreiner2019theorem}, which helps students write and check logical specifications through model checking. Unlike Hazel Deriver, which focuses on interactive editing and step-by-step scaffolding, RISCAL checks properties of logical models over finite domains and does not support derivation tree construction.

Some math education tools, such as the one described by Back\cite{back2010structured}, promote structured editing and reuse of proof components in setting of algebraic proofs. These ideas influence Hazel Deriver's design as well, especially its support for reusable subtrees and modular structure.

The underlying structural editor in Hazel Deriver is part of ongoing research on tile-based editing, as explored in the Tylr system\cite{moon2022tylr}, which investigates how to support linear editing affordances while preserving term structure.
\section{System Overview}

\begin{figure*}[htbp]
\centerline{\includegraphics[width=0.95\textwidth]{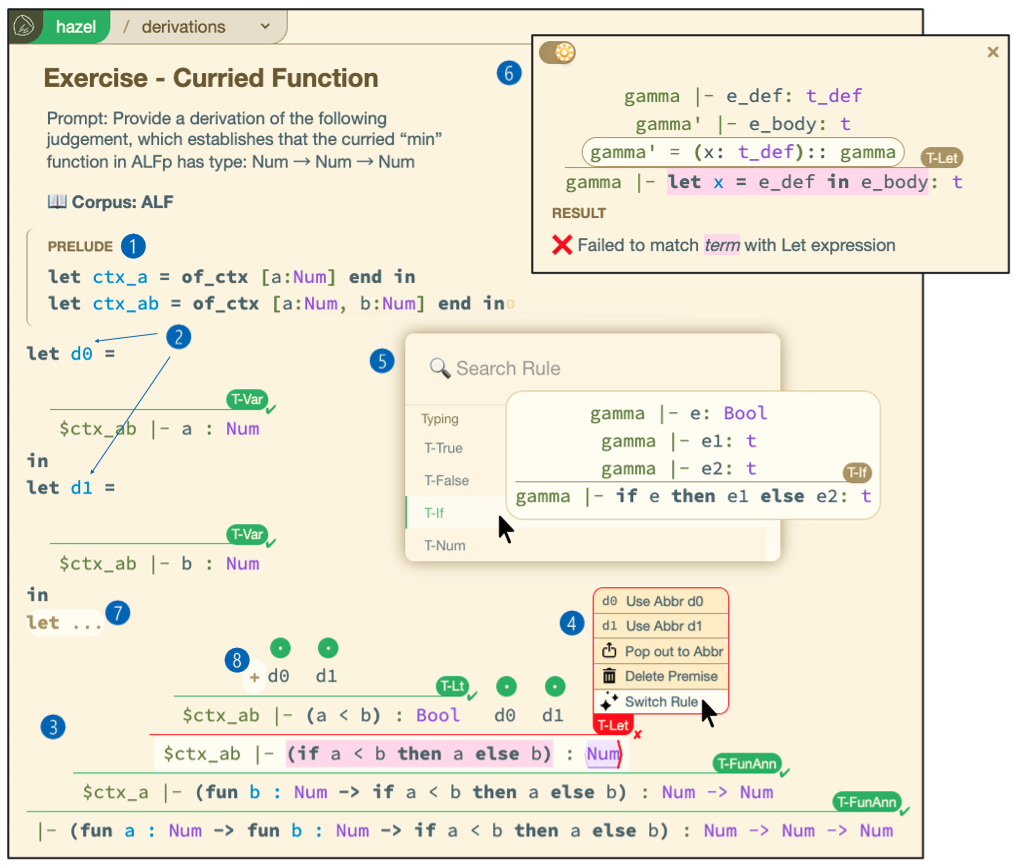}}
\caption{
Hazel Deriver user interface. 
\circledred{1} Prelude: reusable assumptions and named abbreviations for commonly used terms.
\circledred{2} Subtrees: reusable derivation fragments.
\circledred{3} RootTree: the main goal derivation tree being constructed.
\circledred{4} Drop-down panel.
\circledred{5} Rule-selection panel: categorizes and allows searching of rules that can be applied to each node.
\circledred{6} Sidebar: rule documentation synced with current selection.
\circledred{7} Add Subtree button: inserts a subtree into the current derivation.
\circledred{8} Add Premise button: adds a premise in a node.
}
\label{fig-system}
\end{figure*}

\subsection{Interface Layout and Core Components}

The \circledred{1} Prelude is where users define shorthand expressions and assumptions (such as contexts, variables, or other bindings) that can be reused throughout the proof. This reduces redundancy and improves clarity.

The \circledred{2} Subtrees area allows students to define and name small derivation fragments. These reusable subtrees promote modular thinking and reduce repeated manual effort.

The \circledred{3} RootTree is the central area where the main derivation is constructed. Each node represents a logical judgment. Students apply rules to nodes, add premises, and fill in required terms to build out the tree step-by-step. The system ensures that edits are well-structured derivations (with holes) at all times.

\subsection{Guided Editing and Verification}

Hazel Deriver checks student edits continuously. Every time a rule is applied or a term is changed, the system runs a verification pass. This process checks that each rule is used correctly, the terms match the expected form, and any conditions attached to the rule are satisfied. The verification engine classifies each node as either correct, incorrect, or indeterminate. Nodes marked as indeterminate contain holes or incomplete parts, so they cannot be fully verified yet, but the rest of the tree can still be checked.

This behavior supports the use of partial proofs. For example, a student can build the structure of a derivation first, leave some parts unfinished, and return later to complete them. Unlike tools that only verify once the full proof is written, Hazel Deriver helps students learn as they go by catching errors early.

\subsection{Interactive Rule Selection and Documentation}

To apply a rule, students use the \circledred{5} rule-selection panel. The panel is searchable and grouped by category, such as typing rules or evaluation rules. When a rule is selected, the \circledred{4} drop-down panel appears to collect required premises and terms.

At the same time, the \circledred{6} Sidebar displays synchronized contextualized documentation\cite{potter2022contextualized} about the selected rule, including its definition and application details. Terms with errors in the derivation are linked to parts of the rule using color highlights, so students can clearly see how their example fits the general pattern. This documentation is generated from the same rule source, ensuring consistency with the verifier. This feature reduces the need to switch between the tool and outside reference materials.

\subsection{Convenient Structural Manipulation}

To manage repeated structure, users can define reusable fragments using the \circledred{7} Add Subtree button. These appear in the \circledred{4} drop-down panel for reuse in later derivation steps.

The system supports a flexible workflow: students can first scaffold the overall structure with typed holes, insert reusable subtrees, and incrementally fill in derivation details. All edits are constrained to maintain structural validity, supporting accessibility and reducing syntax-level distractions.

\subsection{Support for Multiple Languages}

A key feature of Hazel Deriver is its support for multiple languages. It currently supports propositional logic, evaluation semantics, and type inference for ALFA--a lightweight educational language designed to teach fundamental programming language concepts. ALFA includes rules for the simply typed lambda calculus and can be extended with additional judgments.

The system allows instructors to define new rule systems using OCaml, and switch between them, making Hazel Deriver adaptable for various teaching contexts.

\subsection{Error Feedback and Highlighting}

When an error is detected, Hazel Deriver highlights problematic subterms directly within the tree (as shown in Fig.~\ref{fig-system}). Problematic nodes or subterms are marked with red highlights and small icons. The error message is conveyed through the \circledred{6} Sidebar. For example, if a student tries to use a rule that expects a function but provides a let-expression, the system might say: “Expected a function term, but found a let-expression.” \\\\

These features are part of Hazel Deriver's layered scaffolding approach--including live feedback, contextual documentation, error localization, and reusable structure support--which helps students understand errors without relying on automatic proof search. This design aligns with principles for interactive mathematical modeling in education\cite{neuper2019technologies}, where guidance is given without oversimplifying the task.
\section{Preliminary Evaluation}

To understand how Hazel Deriver supports learning, we conducted a small user study focused on usability, feature impact, and the balance between system guidance and student control. We aimed to observe how students interact with the system during different types of derivation tasks and how they respond to its layered scaffolding features. The full survey, task details, and data summary are included as supplemental material.

\subsection{Study Setup}

The study recruited 7 participants, all of whom had completed a programming languages course or had prior experience with evaluation and typing judgments for the simply typed lambda calculus.

Participants completed a series of derivation tasks commonly found in undergraduate assignments: transcription (copying a written derivation into the tool), debugging (correcting flawed derivations), and construction (building a derivation from a goal judgment). Each task highlighted different system features. We collected behavioral data and post-task survey responses to assess perceived difficulty and tool effectiveness.

\subsection{Study Results}

\begin{figure*}[htbp]
\centering
\begin{subfigure}[t]{0.48\textwidth}
    \includegraphics[width=\linewidth]{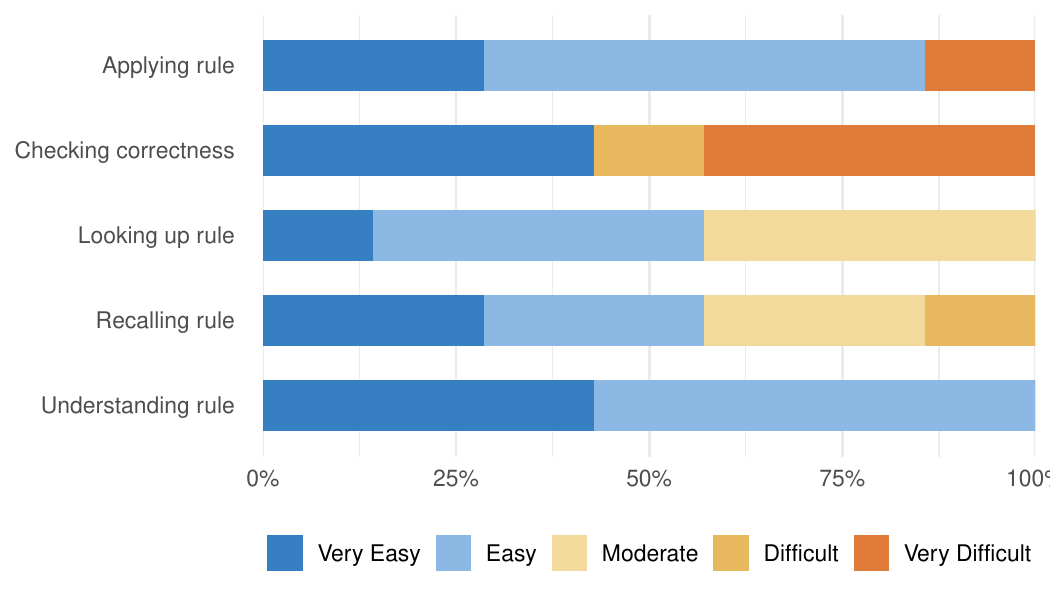}
    \caption{Handwritten derivations}
    \label{fig-diff-hand}
\end{subfigure}
\hfill
\begin{subfigure}[t]{0.48\textwidth}
    \includegraphics[width=\linewidth]{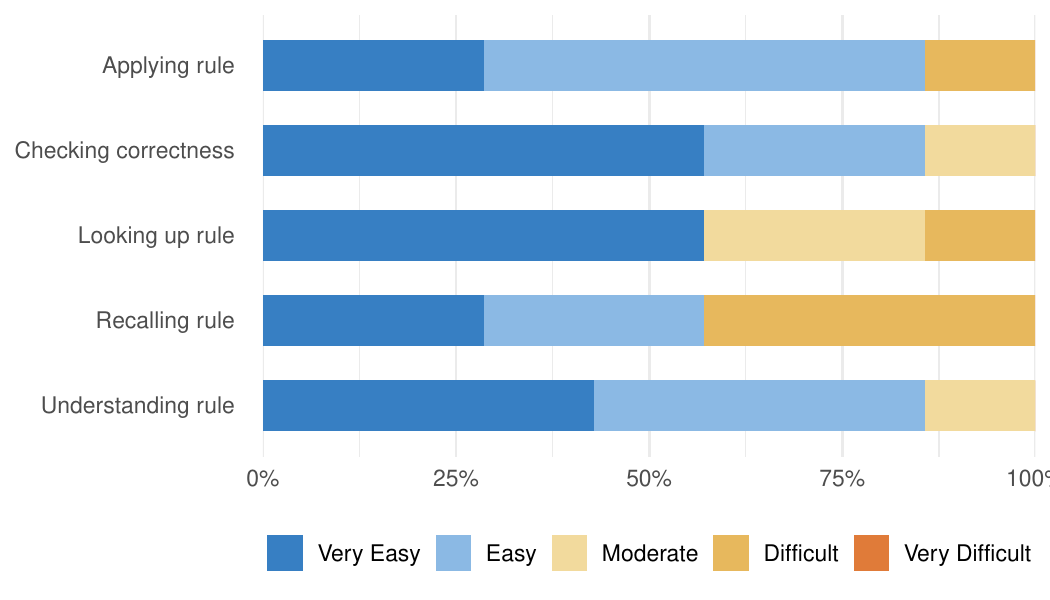}
    \caption{Hazel Deriver}
    \label{fig-diff-deriver}
\end{subfigure}
\caption{Reported difficulty of derivation steps across two settings. Participants rated each step's difficulty when constructing derivations by hand (a) versus using Hazel Deriver (b).}
\label{fig-diff-comparison}
\end{figure*}

\textbf{RQ1} asks how Hazel Deriver affects students' perceived difficulty and confidence during derivation tasks. Our results suggest that its layered scaffolding features significantly lowered the perceived difficulty, particularly in error checking, and boosted students' confidence. Participants described the editing experience as more fluid and forgiving than hand-written derivations, which previously caused frustration and hesitation. All but one participant gave it the highest possible score for helping them identify and correct mistakes early.

Figures~\ref{fig-diff-hand} and~\ref{fig-diff-deriver} show self-reported difficulty across derivation steps for handwritten versus Hazel Deriver workflows. Notably, steps involving error checking saw the largest perceived improvements.

Participants reported that the immediate validation of each edit reduced their anxiety about making mistakes. One student commented, “I used to be afraid of writing something wrong and then having to erase the whole tree. Here, I just try something, and if it's wrong, I know right away.”.
In comparing Hazel Deriver to traditional hand-written derivations, participants described the latter as “stressful,” “messy,” and “easy to mess up.” Most students reported that they previously struggled to identify which rule should be applied, and that writing large trees by hand made revision difficult. In contrast, the structured editor in Hazel Deriver allowed them to iterate more confidently. One student noted: “I used to get lost halfway through. Here, I can see exactly where I went off track.”

The presence of typed holes was cited as a major strength, where students could leave parts of the derivation temporarily incomplete. These holes enabled exploratory derivation, where students would start from a root and fill in what they could while postponing harder subgoals. This behavior suggests that students viewed derivation as a process of refinement, rather than a one-shot construction.

All participants stated that they would have benefited from using Hazel Deriver in their earlier coursework.

\subsection{Observed Patterns}

To answer \textbf{RQ2}, we analyzed usage patterns related to the system's layered scaffolding features: live feedback, rule documentation, structural editing, and error localization.

Students often began by using the rule table but shifted toward relying on sidebar documentation and rule previews after the first task. Several reported that color-linked documentation helped them map rule patterns to their examples more easily.

Live feedback and error highlighting supported iterative construction. For instance, most participants revised individual nodes multiple times without deleting large parts of their tree, a behavior that is harder to achieve in handwritten workflows.

However, some participants expressed concern about their dependence on system feedback. A few students struggled when asked to choose rules without hints. This suggests that while scaffolding supported exploration and reduced error, it may also have suppressed active recall in some contexts.

\subsection{Limitations}

This was a small, exploratory study with 7 participants, all of whom had some background in logic or programming languages. Tasks were completed in a low-pressure environment, not in a real classroom or exam setting. As a result, the findings may not reflect how beginners or larger groups would interact with Hazel Deriver in practice. In classroom settings, students face additional constraints such as time limits, collaboration, and grading policies. We also did not measure long-term learning outcomes, such as how well students retain concepts or apply them in future tasks. Future studies should track performance over time to better understand the system's performance.

\section{Discussion}

Our study of Hazel Deriver reveals both the benefits and the trade-offs of using system scaffolding to support formal reasoning tasks. Students completed derivations more confidently and with fewer errors, but the tool also introduced risks of over-reliance. These findings highlight the importance of adjusting the level of support to match student needs and experience.

\subsection{The Role of Scaffolding in Learning}

Hazel Deriver includes several forms of scaffolding, to help students as they build derivation trees. These supports help students progress more gradually and gain confidence in their work. However, some students said the tool made things feel too easy. They could try different rules until something worked, without fully understanding why. This “trial-and-error” approach can turn derivation into a guessing game instead of a reasoning process. This concern is similar to ideas from educational research\cite{chi2009active}, which warn that students may learn less if tools do too much of the work for them. 

\subsection{Engagement through Iteration and Visualization}

Students said that Hazel Deriver made it easier to revise their work. The interface helped them quickly see what was wrong and try different ideas without starting over. This made students feel more engaged and less anxious. The visual feedback helped them understand their mistakes and see why a particular rule or structure did not work. The tool's use of typed holes also encouraged students to explore and build their derivations piece by piece. Many saw derivation as a process of refinement rather than something they had to get right in one try. It also aligns with research showing that immediate, specific feedback improves learning\cite{hattie2007power}.

\subsection{Balancing Guidance and Independence}

While the system effectively supported students in getting started, it also raised questions about how much support is too much. In some cases, students relied heavily on system features and struggled when asked to work more independently. This reflects a broader challenge in educational tool design: how to support learners without reducing meaningful problem-solving. Prior research on adaptive scaffolding\cite{azevedo2005scaffolding} suggests that the best tools provide support when needed, but gradually reduce help as students gain confidence. Hazel Deriver could incorporate adjustable scaffolding levels so as to offer more help early on, and less later in the course.

\section{Future Work}

Based on our study and user feedback, we have identified several directions to improve Hazel Deriver in future versions.

\textbf{Adjustable Scaffolding Levels. } Some features we are considering such as automatic rule suggestions or semi-automated subtree generation could help students write proofs faster. Although common in similar tools and could improve usability, we need to better understand how they affect learning. While useful in practice, their pedagogical effectiveness remains uncertain and warrants further study.

\textbf{Improved Error Explanations.} Some students felt that the messages they received were hard to interpret. For example, when a rule is applied incorrectly, the system currently highlights the mismatch in the conclusion form. But this can be confusing if the real problem is that the student chose the wrong rule entirely. In future versions, we want to improve these messages so that they guide students more effectively without causing misunderstandings.

We plan to conduct broader studies involving entire course cohorts using Hazel Deriver. These studies will help us see how the tool affects learning over time, including how well students retain knowledge and how their attitudes toward formal logic develop. 


\bibliographystyle{IEEEtran}
\bibliography{references}

\begin{thebibliography}{10}
\providecommand{\url}[1]{#1}
\csname url@samestyle\endcsname
\providecommand{\newblock}{\relax}
\providecommand{\bibinfo}[2]{#2}
\providecommand{\BIBentrySTDinterwordspacing}{\spaceskip=0pt\relax}
\providecommand{\BIBentryALTinterwordstretchfactor}{4}
\providecommand{\BIBentryALTinterwordspacing}{\spaceskip=\fontdimen2\font plus
\BIBentryALTinterwordstretchfactor\fontdimen3\font minus \fontdimen4\font\relax}
\providecommand{\BIBforeignlanguage}[2]{{%
\expandafter\ifx\csname l@#1\endcsname\relax
\typeout{** WARNING: IEEEtran.bst: No hyphenation pattern has been}%
\typeout{** loaded for the language `#1'. Using the pattern for}%
\typeout{** the default language instead.}%
\else
\language=\csname l@#1\endcsname
\fi
#2}}
\providecommand{\BIBdecl}{\relax}
\BIBdecl

\bibitem{bull1983derivation}
C.~Bull and R.~Frydman, ``The derivation and interpretation of the lucas supply function,'' \emph{Journal of money, credit and banking}, vol.~15, no.~1, pp. 82--95, 1983.

\bibitem{omar2019live}
C.~Omar, I.~Voysey, R.~Chugh, and M.~A. Hammer, ``Live functional programming with typed holes,'' \emph{Proceedings of the ACM on Programming Languages}, vol.~3, no. POPL, pp. 1--32, 2019.

\bibitem{potter2022contextualized}
H.~Potter, A.~Madadi, R.~Just, and C.~Omar, ``Contextualized programming language documentation,'' in \emph{Proceedings of the 2022 ACM SIGPLAN International Symposium on New Ideas, New Paradigms, and Reflections on Programming and Software}, 2022, pp. 1--15.

\bibitem{aldrich2008sasylf}
J.~Aldrich, R.~J. Simmons, and K.~Shin, ``Sasylf: An educational proof assistant for language theory,'' in \emph{Proceedings of the 2008 international workshop on Functional and declarative programming in education}, 2008, pp. 31--40.

\bibitem{o2022holbert}
L.~O'Connor and R.~Amjad, ``Holbert: Reading, writing, proving and learning in the browser,'' \emph{arXiv preprint arXiv:2210.11411}, 2022.

\bibitem{keenan2024learner}
M.~Keenan and C.~Omar, ``Learner-centered design criteria for classroom proof assistants,'' \emph{Proceedings of HATRA}, 2024.

\bibitem{eggensperger2013proof}
C.~Eggensperger, ``Proof pad: A new development environment for acl2,'' \emph{arXiv preprint arXiv:1304.7856}, 2013.

\bibitem{staudacher2018conception}
K.~Staudacher, ``Conception, implementation and evaluation of proof editors for learning,'' \emph{Bachelor thesis, Institute of Computer Science, LMU, Munich}, 2018.

\bibitem{mendes2014structure}
A.~Mendes, R.~Backhouse, and J.~F. Ferreira, ``Structure editing of handwritten mathematics: Improving the computer support for the calculational method,'' in \emph{Proceedings of the Ninth ACM International Conference on Interactive Tabletops and Surfaces}, 2014, pp. 139--148.

\bibitem{schreiner2019theorem}
W.~Schreiner, ``Theorem and algorithm checking for courses on logic and formal methods,'' \emph{arXiv preprint arXiv:1904.00620}, 2019.

\bibitem{back2010structured}
R.-J. Back, ``Structured derivations: a unified proof style for teaching mathematics,'' \emph{Formal Aspects of Computing}, vol.~22, pp. 629--661, 2010.

\bibitem{moon2022tylr}
D.~Moon, A.~Blinn, and C.~Omar, ``Tylr: a tiny tile-based structure editor,'' in \emph{Proceedings of the 7th ACM SIGPLAN International Workshop on Type-Driven Development}, 2022, pp. 28--37.

\bibitem{neuper2019technologies}
W.~Neuper, ``Technologies for" complete, transparent \& interactive models of math" in education,'' \emph{arXiv preprint arXiv:1904.08751}, 2019.

\bibitem{chi2009active}
M.~T. Chi, ``Active-constructive-interactive: A conceptual framework for differentiating learning activities,'' \emph{Topics in cognitive science}, vol.~1, no.~1, pp. 73--105, 2009.

\bibitem{hattie2007power}
J.~Hattie and H.~Timperley, ``The power of feedback,'' \emph{Review of educational research}, vol.~77, no.~1, pp. 81--112, 2007.

\bibitem{azevedo2005scaffolding}
R.~Azevedo and A.~F. Hadwin, ``Scaffolding self-regulated learning and metacognition--implications for the design of computer-based scaffolds,'' \emph{Instructional science}, vol.~33, no. 5/6, pp. 367--379, 2005.

\end{thebibliography}

\end{document}